\documentclass[format=acmsmall]{acmart}
\renewcommand\footnotetextcopyrightpermission[1]{} 
\usepackage{listings}
\usepackage{algpseudocode}
\setcopyright{none}
\begin{document}
\title{Fast GPU bounding boxes on tree-structured scenes}
\author{Raph Levien}
\affiliation{
    \institution{Google}
    \city{San Francisco}
    \state{CA}
    \country{USA}
}

\newcommand{\Bic}{\textrm{Bic}}
\newcommand{\Stk}{\textrm{Stk}}

\begin{abstract}
Computation of bounding boxes is a fundamental problem in high performance rendering, as it is an input to visibility culling and binning operations. In a scene description structured as a tree, clip nodes and blend nodes entail intersection and union of bounding boxes, respectively. These are straightforward to compute on the CPU using a sequential algorithm, but an efficient, parallel GPU algorithm is more elusive. This paper presents a fast and practical solution, with a new algorithm for the classic parentheses matching problem at its core. The core algorithm is presented abstractly (in terms of a PRAM abstraction), then with a concrete mapping to the thread, workgroup, and dispatch levels of real GPU hardware. The algorithm is implemented portably using compute shaders, and performance results show a dramatic speedup over a sequential CPU version, and indeed a reasonable fraction of maximum theoretical throughput of the GPU hardware. The immediate motivating application is 2D rendering, but the algorithms generalize to other domains, and the core parentheses matching problem has other applications including parsing.
\end{abstract}

\maketitle
\thispagestyle{empty}
\pagestyle{plain}

\section{Introduction}

This paper presents an efficient parallel solution to the computation of bounding boxes on a scene structured as a tree. The data model for the tree includes both a ``clip node,'' which is a shape intersected with all its children, and a ``blend node,'' which represents some compositing operation applied to its children. While the immediate application is 2D rendering, this data model is similar to constructive solid geometry, where the nodes might represent intersection and union, respectively. The goal of the computation is to compute bounding boxes on all nodes in the tree, based on the bounding boxes of individual elements as well as the tree structure. The bounding box of a leaf node is that object's bounding box intersected with the intersection of all clip nodes on the path to the root, and the bounding box of a blend node is the union of all leaf nodes which are descendants of that node.

A sequential algorithm for this task is straightforward: walk the tree and maintain a \emph{stack,} pushing an entry every time traversing from a parent to a child, and popping when traversing from child to parent. Each element of the stack contains two bounding boxes, one for clipping and one for blending, with straightforward logic for computing bounding box intersection on each clip node and union on each blend.

Adapting this sequential algorithm to efficient parallel computation on GPU is challenging for two primary reasons. First, there are complex data dependencies from one node in the tree to another. Second, the stack is a variable-sized object and thus cannot directly be used as the element of a scan (generalized prefix sum) operation; if the stack were bounded to a very shallow depth, then scan algorithms could be practically employed.

This paper presents a fully parallel, high performance solution to the problem, without limitations on tree depth. At its heart is a practical solution to the \emph{parentheses matching problem.} We develop two versions of this core algorithm, with different practical performance characteristics. The simpler algorithm has a work factor proportional to the size of a GPU workgroup, but achieves good utilization over a range of problem sizes. Here, work factor is defined as the total number of steps times the number of processors, divided by the number of steps for a sequential implementation. A more complex version is based on a \emph{work efficient} abstract algorithm and achieves higher peak throughput, but poorer utilization for smaller problem sizes. Based on the results of this empirical evaluation, we develop the full bounding box solution on top of the simpler algorithm. For applications for which extremely large problem sizes are anticipated, it is possible to adapt the work-efficient algorithm. The 2D renderer, piet-gpu, uses stream compaction, another well known application of prefix sums, to extract \emph{just} the clip and blend nodes from the scene description, as opposed to having to run the algorithm on all nodes in the source scene description.

While the immediate application is computation of bounding boxes, which is important and fundamental to a variety of rendering tasks, the algorithm is stated in terms of general associative operations with identity (monoids). For example, a potential application is applying 3D transformations, as in relative orientations of bones in a rig, to a skeleton represented as a tree structure with abitrary size and depth. In addition, the parentheses matching problem itself has significant other applications including parsing, bin packing\cite{And89} and tree pattern matching\cite{Pla20}.

\subsection{Motivation}

In many rendering applications, the most natural representation for the scene is a tree, which is more general than the usual large flat array of values. However, parallel processing of the tree structure on the GPU is challenging, and often much of the rendering task is therefore relegated to the CPU. A promising approach for efficient parallel processing of tree-structured data is a \emph{flattened representation} in which a node with children is encoded as an open parenthesis, the encoding of its children, then a close parenthesis. Given this representation, the parenthesis matching problem produces a representation of the tree structure which can then be used to solve other problems such as bounding box computation. Theoretical algorithms for parentheses matching date back to the '80s but practical implementations are sparse; the only one known to the author is \cite{Voe21}.

The computation of bounding boxes is motivated by a desire to improve the performance of the open-source piet-gpu 2D rendering engine. A previous version required computation of bounding boxes on the CPU. The renderer performs most other rendering tasks on the GPU, using a pipeline of compute shaders, so moving bounding box computation to the GPU was a special concern. That led to exploration of whether known techniques for GPU processing of tree-structured data could be adapted to the problem of computing bounding boxes. Fortunately, such an approach does appear to be practical, and this paper describes an implementation in detail as well as characterizing its performance on a diverse selection of real GPU hardware.

\subsection{Limitations of current state of the art}

The literature on the parentheses matching problem goes back decades (at least to \cite{Bar85}), but until now there is no known satisfactory solution running on actual GPU hardware. The literature falls into several categories:

\begin{itemize}
\item Theoretical presentations of work-efficient algorithms analyzed in terms of an abstract Parallel Random Access Machine PRAM) model but no clear mapping to an efficient GPU implementation (\cite{Bar85}, \cite{Lev92}, \cite{Pra94}).

\item Practical algorithms which run on GPU but have a work factor dependent on maximum nesting depth (\cite{Hsu19}).

\item More limited GPU-based parsing algorithms which cannot handle arbitrary tree structure (\cite{Ste19}). This category also includes the use of standard generalized prefix sum algorithms with a small fixed bound on nesting depth.
\end{itemize}

Thus, the prevailing wisdom remains that parsing of arbitrary tree structured data is inherently a serial problem and must be done on CPU rather than GPU.

\subsection{Key insights and contributions}

There are several key insights in this paper, culminating in presentation and empirical performance measurement of an algorithm that is fast and practical to implement on standard GPU hardware.

The first insight is that the parentheses matching problem can be expressed in terms of two monoids, both of which can be used to compute matches, but with different time/space tradeoffs. The first of these is the bicyclic semigroup\cite{Cli67} which is cheap to compute and for which a parenthesis match can be computed using binary search, and the second is a ``stack monoid'' which takes more space but gives a parenthesis match in $O(1)$ time. Either by itself can be used to derive an algorithm which is parallel but has $O(\log n)$ work factor.

The second insight is that \emph{interleaving} these two approaches yields a work-efficient algorithm. Further, the two approaches map well to the hierarchical structure of actual GPU hardware, in which each GPU workgroup computes one partition of the input problem. We present a simple algorithm consisting of reduction of the stack monoid (computing one stack snapshot at the beginning each partition), followed by binary search of the bicyclic semigroup to resolve matches within a partition. The second step can be done within a workgroup, using efficient shared memory. It is reasonably fast but not work-efficient.

A more complex version of the algorithm adds a third level of hierarchy: a sequence of $k$ elements processed per thread, instead of just one as in the simpler algorithm. This technique is analogous to that used for high performance prefix sum implementations, but requires more sophistication. Stack monoid reduction is used for the smallest granularity, then binary search for the workgroup level, and then stack monoids again for finding matches across workgroup boundaries. This algorithm produces greater peak throughput, but because the run time of an individual thread is considerably longer, it is also more challenging to achieve good utilization.

Given the tree structure as computed by parentheses matching, we also present a efficient parallel algorithm for computing intersections and unions of bounding boxes, flowing data down and up the tree, respectively.

\subsection{Experimental methodology and artifact availability}

The primary empirical claim is that the proposed algorithm is fast on standard GPU hardware. To demonstrate this claim, we run the code on an AMD 5700 XT as Vulkan compute shaders. The test consists of a random sequence of parenthesis. The GPU time is measured with Vulkan and Metal timer queries.

All software is available on GitHub with a permissive Apache 2 open source license\footnote{\texttt{https://github.com/linebender/piet-gpu}, particularly the \texttt{stack} test in the \texttt{tests} subdirectory}. The infrastructure for running and measuring compute shader performance is cross-platform and runs on Metal and Direct3D 12 as well as Vulkan.

\section{The parentheses matching problem}

The classical version of the parentheses matching problem is, for every index in the source string, find the index of the corresponding matching parenthesis. This paper actually considers a stronger version of the problem: for every closing parenthesis, find the index of the matching open parenthesis. But for every opening parenthesis, find the index of the immediately enclosing opening parenthesis. It is straightforward to reconstruct the traditional version, but the converse is not true.

One statement of the problem is as a simple sequential program which uses a stack, as shown in Figure~\ref{fig:sequential}.

\begin{figure}
\begin{lstlisting}
stack = [-1]
for i in range(len(s)):
    out[i] = stack[len(stack) - 1]
    if inp[i] == '(':
        stack.push(i)
    elif inp[i] == ')':
        stack.pop()
\end{lstlisting}
\caption{Sequential algorithm for parentheses matching}
\label{fig:sequential}
\end{figure}

We will compute as intermediate results \emph{snapshots} of the stack at step $i$, in other words the value of the \texttt{stack} variable during iteration of the sequential algorithm. An appealing quality of this specific formulation of the parentheses-matching problem is that all stacks can be recovered from the output, just by repeatedly following references until the root is reached (here represented by a value of -1).

\section{The bicyclic semigroup}

The theoretical derivation of the algorithm relies heavily on the \textit{bicyclic semigroup,} actually a monoid, which is known to model the balancing of parentheses. An element of the bicyclic semigroup $\Bic$ can be represented as a pair of nonnegative integers, with $(0, 0)$ as an identity and the following associative operator (associativity and other properties are given in \cite{Cli67}):

\[
    (a, b) \oplus (c, d) = (a + c - \min(b, c), b + d - \min(b, c))
\]

An open parenthesis maps to $(0, 1)$ and a close parenthesis maps to $(1, 0)$. We will overload the function $\Bic(s)$ over a string to result in the $\oplus$-reduction of this mapping applied to the elements of the string; thus $\Bic(\texttt{\textquotesingle))()(\textquotesingle}) = (2, 1)$. We will use slice notation on strings; $s[i..j]$ represents the substring beginning at index $i$ of length $j - i$.

The bicyclic semigroup gives rise to an alternate definition of the parentheses matching problem. In particular, parenmatch(s)[j] is the maximum value of $i$ such that $\Bic(s[i..j]).b = 1$. This is one less than the minimum value such that the $b$ field is 0. Note that $\Bic(s[i..j]).b$ is monotonically increasing as $i$ decreases.

\section{The stack monoid}

Another related monoid is the \emph{stack monoid,} which is a sort of hybrid of the bicyclic semigroup and the free monoid (which can be defined as sequences of arbitrary values). Essentially, rather than just counting the number of stack pushes, it contains the actual values pushed on the stack.

Like the bicyclic semigroup, the stack monoid can be represented as a 2-tuple. The first element in the tuple is the number of unmatched closing parentheses, the same as the bicyclic semigroup. The second element is a \emph{sequence} of values corresponding to unmatched open parentheses, as opposed merely to their count as in the bicyclic semigroup. In the context of this paper, those values are typically the indices, though the monoid is free in that it can be defined over any sequence element type.

The empty stack monoid is $(0, [])$. The value corresponding to an open parenthesis with associated value $x$ is $(0, [x])$, and the value corresponding to a close parenthesis is $(1, [])$. The combination rule is as below:

\[
    (a_0, l_0) \oplus (a_1, l_1) = (a_0 + a_1 - \min(|l_0|, a_1), l_0[..\max(0, |l_0| - a_1)] + l_1)
\]

Like the bicyclic semigroup, the stack monoid lends itself to a straightforward definition of the parenthesis matching problem. A reduction of the stack monoid over a prefix of the input represents a snapshot of the stack, as computed by the sequential algorithm, up to the end of that slice. The result of the parentheses matching algorithm is then the top of the stack at each step.

The parenthesis match value at $j$ is the topmost value of stack snapshot taken at position $j$. Here we use $enum(s)$ to represent the enumeration of the indices of the sequence $s$, for example, $enum(\texttt{\textquotesingle))(\textquotesingle})$ is the sequence $[(0, \texttt{\textquotesingle)\textquotesingle}), (1, \texttt{\textquotesingle)\textquotesingle}), (2, \texttt{\textquotesingle(\textquotesingle})]$.

\[
    parenmatch(s)[j] = last(\Stk(enum(s)[..j]))
\]

The $k$-suffix of the stack monoid is simply the last $k$ values.

The storage required by a single stack monoid value is unbounded, but that does not preclude efficient implementations. In particular, the combination of two values of size $k$ can be done in-place by $2k$ processors in one step. This result generalizes to combination of a vector of values, which can be represented as a stream compaction.

When a sequence containing only unbalanced close parentheses (and no unbalanced open parentheses) is appended to a first sequence, the resulting stack monoid is a prefix of that of the first sequence. Stated more formally:

\begin{multline*}
    rev(\Stk(enum(s)[i_0..i_2]))[k] = rev(\Stk(enum(s)[i_0..i_1]))[k + j] \\
    \textrm{where}\ \Bic(s[i_1..i_2]) = (j, 0)
\end{multline*}

The significance of this relation is that, given the value of the $\Bic$ monoid and a materialized stack slice, it is possible to compute a parenthesis match in $O(1)$ time, as a simple lookup. In cases where $\Bic([s_1..i_2])$ has a nonzero $.b$, the match is found within the slice $s[i_1..i_2]$; a general matching algorithm will look up the match inside the slice if $.b \neq 0$, and use $.a$ to index into a stack monoid value in $s[i_0..i_1]$ when $.b = 0$.

\section{Core parentheses matching parallel algorithm}

The core parallel algorithm is a binary search over the bicyclic semigroup. That algorithm by itself is fully parallel and reasonably efficient; it has a work factor of $O(\log n)$ for the binary search. This algorithm is based on \cite{Bar85} with some refinement; the separate parentheses balance computation and MIN-LEFT/MIN-RIGHT phases of that algorithm are combined into a single monoid (the bicyclic semigroup), which is both faster since it's a single pass and also more easily generalizable to other monoids.

As an initial step in running this algorithm, a binary tree of bicyclic semigroup values is constructed; this is the same as the up-sweep phase of a standard parallel prefix sum implementation\cite{Ble90}. Specifically, the leaf nodes of the tree are defined by $tree[0][i] = \Bic(s[i])$, and parent nodes by the relation $tree[j+1][i] = tree[j][2i] \oplus tree[j][2i+1]$. Construction of this tree takes $\lg n$ steps, and the tree itself requires storage of $2n - 2$ bicyclic semigroup elements.

Then, for each index $i$, the algorithm shown in Figure~\ref{fig:core} searches the tree for a parentheses match.

\begin{figure}
\begin{algorithmic}
    \State{$i \gets i_1$}
    \State{$p \gets (0, 0)$}
    \State $j \gets 0$
    \While{$j < \lg w$}
        \If{$i\; \textrm{bitand}\; 2^j \neq 0$}
            \State $q \gets tree[j][\lfloor i/2^j\rfloor - 1] \oplus p$
            \If{$q.b = 0$}
                \State{$p \gets q$}
                \State{$i \gets i - 2^j$}
            \Else
                \State{\textbf{break}}
            \EndIf
        \EndIf
        \State $j \gets j+1$
    \EndWhile
    \If{$i>0$}
        \While{$j > 0$}
            \State $j \gets j-1$
            \State $q \gets tree[j][\lfloor i/2^j\rfloor - 1] \oplus p$
            \If{$q.b = 0$}
                \State{$p \gets q$}
                \State{$i \gets i - 2^j$}
            \EndIf
        \EndWhile
    \EndIf
\end{algorithmic}
\caption{Core parallel matching algorithm}
\label{fig:core}
\end{figure}

On termination, $i$ contains the smallest value such that $\Bic(s[i..i_1]).b = 0$, thus $i - 1$ is the solution to the parentheses matching problem.

\begin{figure*}
    \includegraphics[width=350pt]{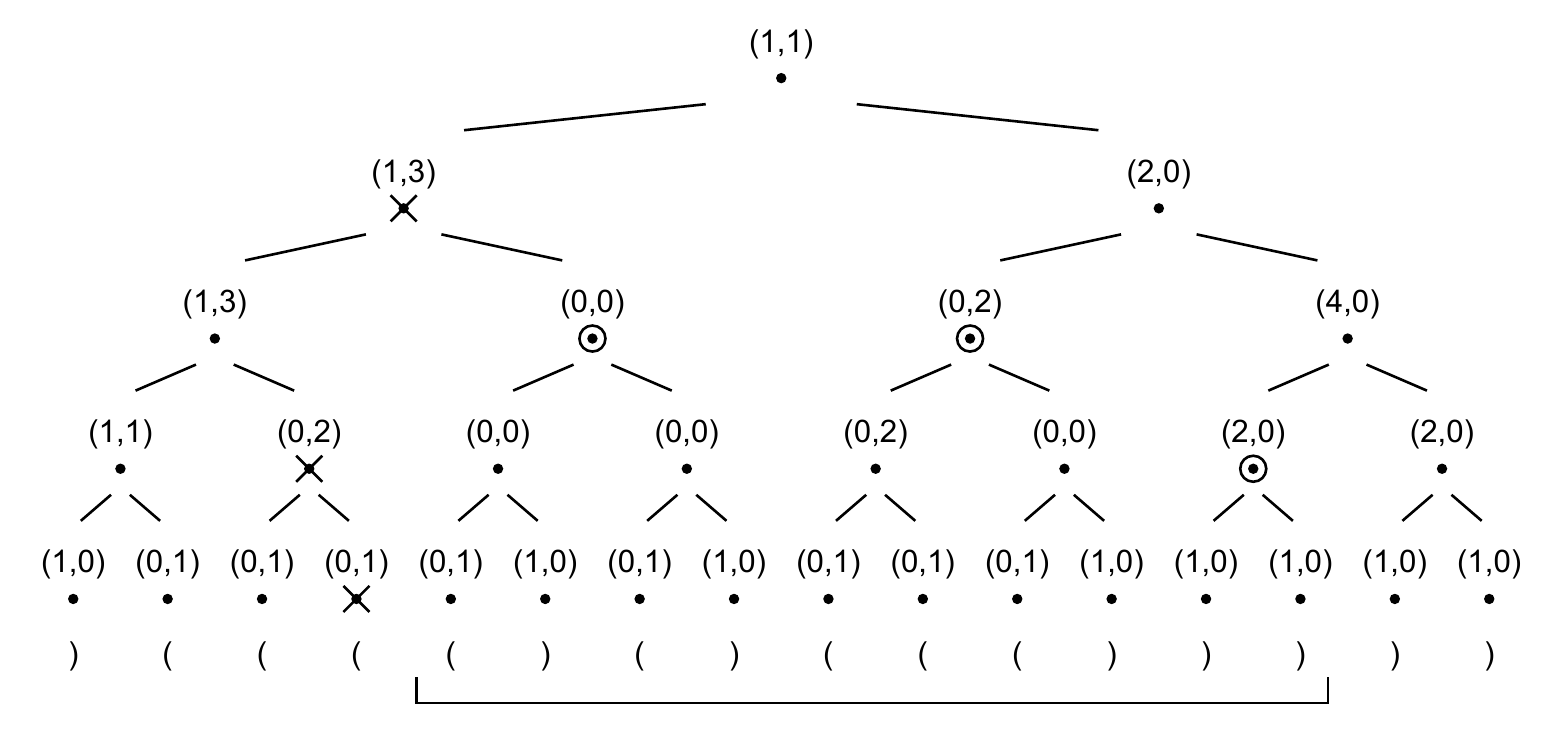}
    \caption{Binary tree search for matching parentheses}
    \label{fig:binarytree}
\end{figure*}

Operation of the algorithm is illustrated in Figure~\ref{fig:binarytree}. Here, $i_1$ is 14 (of a 16 element sequence), and the final value of $i$ is 4, indicating that $\Bic(s[4..14]).b = 0$ but $\Bic(s[3..14]).b = 1$. There is an upward scanning pass followed by a downward scanning pass. At each level, one node from the tree is examined. If combining that node with $b$ would preserve $.b = 0$, it is incorporated (and $i$ adjusted to point to the beginning of the range covered by the node), otherwise it is rejected. Nodes incorporated are marked with a circle, nodes rejected by an X.

This binary search takes $2\lg n$ steps in the worst case. Thus, while the algorithm is highly parallel, it is not considered work-efficient in a theoretical sense.

\section{Parallel bounding boxes}

With the tree structure fully computed as a parenthesis match value for every element in the input, we can now solve the bounding box problem.

First, the data dependencies between the clip and blend components of the problem suggest a two-pass approach. In the first pass, clips are computed, assigning each leaf node a clipped bounding box. In the second pass, blends are computed as a union of clipped bounding boxes. Both passes require the tree structure and use parentheses matching, but otherwise the details are different.

\subsection{Clips; intersections}
\label{section:clip}

The definition of the clip problem is: each leaf node is assigned the intersection of its inherent bounding box and the clip nodes above it in the path to the root. This is actually fairly similar to the definition of (inclusive) prefix sum, where each element is assigned the sum of its own value and the sum of all values preceding it. The difference is in the ``precedes'' relation. In the case of prefix sum, it is a simple comparison of array indices. In the case of clipping, it is a parent relationship in the tree.

Fortunately, the algorithm can be adapted. Each node (represented as an open parenthesis in the flattened representation) has an associated \emph{link} field, which is initially just the parent for each open parenthesis and null otherwise. On each the algorithm updates the bounding box and link state according to the following rules:

\begin{algorithmic}
    \If{$link[i] \neq null$}
        \State{$bbox[i] \gets intersect(bbox[i], bbox[link[i]])$}
        \State{$link[i] \gets link[link[i]]$}
    \EndIf
\end{algorithmic}

One simple way to understand this algorithm is that before step $i$, the bounding box is the intersection of a chain of $2^i$ parents, and link points to the ancestor $2^i$ levels up the tree. After the step, the two chains each of length $2^i$ are combined, resulting in a chain of length $2^{i+1}$.

\subsection{Blends; unions}

Blends are upward flow of data in a way that clips are downward, so the problems are duals in a way, but the algorithm is different. Unlike the clip case where the parent relationship results in a linear chain, blends require merging of data, requiring a different approach.

Fortunately, ``all descdendant nodes'' is relatively easy to express in a flattened tree representation; it's a consecutive slice of nodes from the first descendant to the last, or in other words all array indices from that of the open parenthesis of the blend node to the close. After running the parentheses matching algorithm, both values are available at the close parenthesis for the blend group.


Taking the union of bounding boxes in a contiguous slice is also a straightforward algorithm that can be computed in parallel in $O(\log n)$. First, a balanced binary tree of bounding box unions is constructed, following exactly the same structure as the construction of the \Bic\ values. Second, the start-to-end union is computed from those values, at most $2\lg n$ operations. In fact, the same logic can be used as the algorithm in Figure~\ref{fig:core}, and in fact these passes can be fused.

\section{Simple GPU parentheses matching algorithm}

In this section, we describe a simple adaptation of the parentheses matching algorithm to GPU which is not strictly work-efficient, but is practical in many circumstances, especially if the problem is small or if the costs associated with code complexity are significant. For simplicity, it is presented as two dispatches, effective up to a problem size of $w^2$, where $w$ is the size of a workgroup (a typical value for workgroup size is 1024 on standard GPU hardware).

\subsection{Stack slices}

The first dispatch computes slices of the stack, with each workgroup computing a partition of $w$ values. More precisely, each workgroup computes $\Stk(enum(s)[p..p+w])$, where $p$ is the start of the partition, in this case $w \cdot i$.

This dispatch is very simple. We do a partition-wide reverse scan of the bicyclic semigroup on the mapping of the input elements, followed by a simple stream compaction step: the index is written if the $.a$ of the scan of all following elements is zero, and the memory location to write is derived from the $.b$ value of that scan.

In more detail, for each index $i$ covering the input, index $p + i$ is written to the output at location $\Bic(s[p..p+w]).b - \Bic(s[p+i..p+w]).b$ if the element is an open parenthesis and $\Bic(s[p+i+1..p+w]).a = 0$. For example, in the sequence \texttt{\textquotesingle)(()(\textquotesingle}, the result of the reverse scan of the bicyclic semigroup is [(1, 2), (0, 2), (0, 1) (1, 1), (0, 1)], and the predicate is true at indices 1 and 4, representing the two unmatched open parentheses. The $.b$ values at these indices are 2 and 1, respectively, resulting in 1 being written to output slot 0 and 4 being written to output slot 1.

In this simpler variant, each thread handles one element, and a simple Hillis-Steele scan\cite{HS86} is used to compute the scan of the bicyclic semigroup.

\subsection{Main matching pass}

The second dispatch performs the main parentheses-matching task, resolving all matches within the partition, and also using the stack slices generated by the previous dispatch for the rest. Each workgroup handles one partition independently, performing the following steps sequentially (separated by workgroup barriers).

\begin{itemize}

\item Materialize the stack for the prefix of the input up to the current partition. This results in the $w$-suffix of $\Stk(enum(s)[..p])$ in workgroup-shared memory. It consists of a reverse Hillis-Steele scan of the bicyclic semigroups produced in the previous step (up to $i$), followed by stream compaction which is a per-element binary search of the $.b$ values for the stack value.

\item Compute a binary tree of the bicyclic semigroup from the elements in the partition. This is a simple up-sweep as described by \cite{Ble90}. This binary tree requires storage of $2w-1$ bicyclic semigroup elements in shared memory storage, and $\lg w$ steps.

\item For each element $j$, find the least value $j$ such that $\Bic(s[p + i..p + j]).b = 0$, searching the binary tree in an upwards then a downwards pass. The algorithm is very similar to that given in \cite{Bar85}.

\item If $i > 0$ then the match is found within the partition, and $p + i - 1$ is written to the output. If $i = 0$ then the match is in outside the partition, and the $\Bic(s[p..p + j]).a$ is used to index into the stack as materialized in the first step.

\end{itemize}

\section{Work-efficient parentheses matching algorithm}

In a PRAM model, a simple Hillis-Steele scan over $n$ elements consists of $n$ processors running $\lceil \lg n \rceil$ steps. Thus, it has a \emph{work factor} of $\lceil \lg n \rceil$ compared to the sequential algorithm running in $O(n)$ steps on one processor.

There are a number of work-efficient variations of the basic Hillis-Steele scan. The most straightforward to implement on GPU is for each thread to process $k$ elements sequentially, amortizing the logarithmic cost over these $k$ elements. In a PRAM model, $n / k$ processors each take $O(\lg (n/k))$ steps, which is work-efficient when $k \geq log n$. See \cite{Har07} and \cite{Sen08} for more discussion of efficient GPU implementation of scan.

\subsection{Work-efficient stack slices}

The work-efficient version of the algorithm for producing stack slices is straightforward, and based on standard techniques. We will present it in a bit of detail, as other parts of the algorithm will use similar techniques.

Recall that production of a stack slice is a stream compaction based on a reverse scan of the bicyclic semigroup. The standard work-efficient algorithm for scan is for each thread to process $k$ elements; this way the cost of the Hillis-Steele scan is amortized over $k$. On an actual GPU, each workgroup will have $w$ threads, so will end up processing $wk$ elements. An argument for the optimality of that approach on an EREW (exclusive read, exclusive write) PRAM is given at the end of section 1.2 of \cite{Ble90}.

Applying that technique, the first step is for each thread to do a sequential reduction of the bicyclic semigroup for $k$ elements. Then a standard (reverse) Hillis-Steele reduction over the resulting $w$ elements, which takes $\lg w$ steps. Lastly, each thread does a sequential walk (also in reverse), starting with the exclusive scan value. At each step, the value is written if the $.a$ field of the bicyclic semigroup is $0$, and the location is determined from the $.b$ field.

The change to the shader code compared to the $k=1$ case is modest, and the speedup is significant, contributing to a speedup of over 2.5 for peak throughput (see section~\ref{section:performance}).

\subsection{Work-efficient matching}

The work-efficient matching algorithm also processes $k$ elements per thread. Significant attention to detail is required to ensure that sequential iteration over each of these $k$ elements is $O(1)$, in addition to tree build and tree search stages which are $O(\log w)$.

The key steps of the algorithm are outlined as follows (with the reader referred to the commented source code for a more complete presentation):

\begin{enumerate}

    \item Reduction of the stack monoid for the prefix preceding this partition. This computes the same stack monoid reduction as in the $k=1$ case, and is also a stream compaction. In the input stage, each thread processes $k$ input stack monoids, resulting in a \emph{segment} of subsequences. A bitmap records for each input whether it contributed any unmatched open parentheses. The segments with nonzero bitmaps are placed in a linked list data structure (computed as scan of the max operation). Next, a reverse scan of the bicyclic semigroups, one value per segment. In the output stage, each thread is resposible for generating $k$ values. The first value is determined by binary search of the scan result, and successive values are determined by iteration: walking backward through the subsequence from the input monoid until the beginning, then finding the previous nonempty subsequence from the same segment (using the bitmap to identify nonempty subsequences), then following the linked list structure. Each of these queries is $O(1)$.

    \item Building a binary tree of bicyclic semigroup. This the same as the $k=1$ case, except that each value is a reduction over $k$ input elements. This stage also builds a bitmap for each sequence of $k$ input elements, with a 1 bit for each unmatched open parenthesis in the sequence.

    \item Two binary searches of that tree. The first begins at the first element in the sequence of $k$ input elements processed by the thread. The second begins at the first unmatched open parenthesis in that sequence. The results of these latter searches induces a linked list structure; walking them backwards reconstructs all stack snapshots at multiples of $k$ elements.

    \item Production of parentheses match output. This is a sequential iteration over $k$ elements as well. Maintain an index which might reference a location in the partition, or an index into the prefix stack monoid (the code uses negative values to represent the latter case). This index is initially the result of the first binary search. Also maintain a local stack, initially empty. At each step, if the stack is nonempty, output that value. If empty, use the index to resolve a match value (reading from the prefix stack monoid if outside the partition). Then process the input element. If an open parenthesis, push its index onto the stack. If a close parenthesis, use the hierarchy of previous computations to resolve the next element: if there are nonzero bits remaining in the bitmap (as computed in step 2), use those; otherwise, if the index is in the partition, follow the linked list link as computed in step 3, and lastly, if outside the partition, decrement the index so it references the next element down in the prefix stack monoid.
\end{enumerate}

\section{GPU implementation of clipping and blending}

As with parentheses matching, the basic algorithms for clipping and blending can be adapted to GPU. As in the pure parentheses matching algorithm, there are two dispatches. The first dispatch computes a stack slice, including bounding boxes as well as indices of the corresponding open parentheses. The second dispatch is very much like the abstract PRAM version, running over a partition of one workgroup, and referencing stack slice values computed by the first dispatch to resolve matches falling outside that partition.

To understand the additional steps for the first dispatch, consider the sequential algorithm. When pushing an open parenthesis, compute the intersection of the bounding box associated with that parenthesis (clip node) with the current top of stack, and push that so that it becomes the new top of stack. Otherwise the algorithm is unchanged. It should be clear then that the bounding boxes in the resulting stack slice are the inclusive scan of bounding boxes in the input, over the bounding box intersection monoid. A partition-wide scan is straightforward to implement in a GPU compute shader.

The second dispatch again follows the same structure as the pure parentheses matching case, with additional logic for bounding boxes. The first stage is reading and combining the partition-wide stack slices as produced by the first dispatch. A scan of \Bic\ values read from the input results in stack slices that are represented in the stack snapshot at the beginning of the current partition. Each nonempty slice contributes a ``top of stack'' bounding box value, which, thanks to the scan of the previous dispatch, is the intersection of the input bounding boxes in that slice. An exclusive scan of those per-partition bounding boxes, combined (intersected) with the values read from the slices, fully reconstructs the stack snapshot, where each resulting element is the intersection of the that element's input bounding box and that of all its parents.

Note also that if the stack depth is limited to the partition size, a simpler approach can be used: read in all elements as in the pure parentheses matching case, then do a simple intersection scan over the resulting reconstructed stack. However, the version with essentially unbounded stack depth is only a bit more complicated and should have roughly the same performance.

At that point, after parallel computation of the intra-workgroup parentheses matching problem, the parallel algorithm presented in Section~\ref{section:clip} for bounding box intersection is run within the workgroup, with one modification. Instead of computation simply terminating when the parent link references outside the workgroup (this would be the root of the tree), it is resolved to an entry in the stack, and that value is intersected with the intra-workgroup value.

The strategy for union is very similar. The scans over the stack slices in the first dispatch and the stack reconstruction stage of the second dispatch are run in reverse, as union data flows \emph{up} the tree rather than down it. Similarly, the intra-workgroup algorithm is run, and for close parentheses that match open parentheses outside the workgroup, the \Bic value is used to select a stack value that includes the union of all bounding boxes back to that corresponding open parenthesis, which is then combined with the intra-workgroup value.

The union algorithm produces a result at the close parenthesis element of the flattened representation. For further processing it might be more useful to store it at the array entry corresponding to the open parenthesis. This can be done with a random-access memory write, and the index of the open parenthesis is readily available as the result of the parenthesis matching step.

\section{Portable compute shaders}

A goal of this work was to develop an algorithm that could be run efficiently and reliably on a wide range of GPU hardware. To this end, we avoided constructs that would pose problems, such as inter-workgroup communication. We also implemented the algorithm on the piet-gpu-hal infrastructure, which runs compute shaders on multiple back-ends, currently Vulkan, Metal, and Direct3D 12.

Our portability layer provides an abstraction over the multiple APIs available for submitting compute jobs to the GPU. The compute shaders are written in GLSL, and compiled to SPIR-V intermediate representation for the Vulkan back-end, as well as translation to HLSL and Metal Shading Language using the standard open-source \texttt{spirv-cross} tool for use on Direct3D 12 and Metal, respectively. All shader translation is done ahead of time, for minimal runtime overhead at execution time. The test executable is approximately 1.5 megabytes and requires no additional runtime dependencies aside from the GPU interface provided by the operating system.

The infrastructure allows for runtime query of specific GPU capabilities beyond a baseline set, so different permutations of shaders can be selected for performance or compatibility reasons. Even so, the final implementation of this algorithm was written with compatibility in mind, using workgroup shared memory for communication between threads, and not relying on subgroups or memory barriers, both of which can pose portability challenges.

\section{Performance results}
\label{section:performance}

In this section we present and analyze performance results for both parentheses matching in isolation, and also the full bounding box computation, on a representative variety of GPU hardware, as well as a comparison to a scalar CPU implementation. For parentheses matching, we examine the tradeoff between the simpler $k=1$ solution and the more complex, theoretically work-efficient version, showing the latter only superior for large problem sizes.

In all cases, the input problem is randomly generated, with ``push'' and ``pop'' operations equally probable unless it would cause a stack underflow; the resulting stack depth is roughly proportional to the square root of the problem size. These implementations are insensitive to stack depth (measured performance with limited stack depth is within measurement error of the data presented).

For GPU measurements, the timings presented are from timer queries provided by the GPU infrastructure. Actually launching a sequence of dispatches incurs additional overhead, but the assumption here is that this overhead would be amortized with additional processing either upstream or downstream of the bounding box computation. The CPU measurements are simple elapsed time.

Where possible, the clock frequency of the GPU was stabilized for performance measurement purposes. On the AMD Radeon 5700 XT, the clock frequency was manually set to very near 2GHz using manufacturer-provided software. On the Nvidia GTX 1080, the ``prefer performance'' setting was enabled in the manufacturer-provided control panel. No comparable setting was needed on the Apple M1 Max or Intel HD 630, as performance was consistent even without explicit setting. The Vulkan graphics API was used in most cases, with the exception of Apple hardware which used Metal. Our infrastructure layer supports Direct3D 12 as well; performance results are essentially identical to Vulkan, so are not presented here separately.

\subsection{Parentheses matching performance}

\begin{figure}
    \includegraphics[width=190pt]{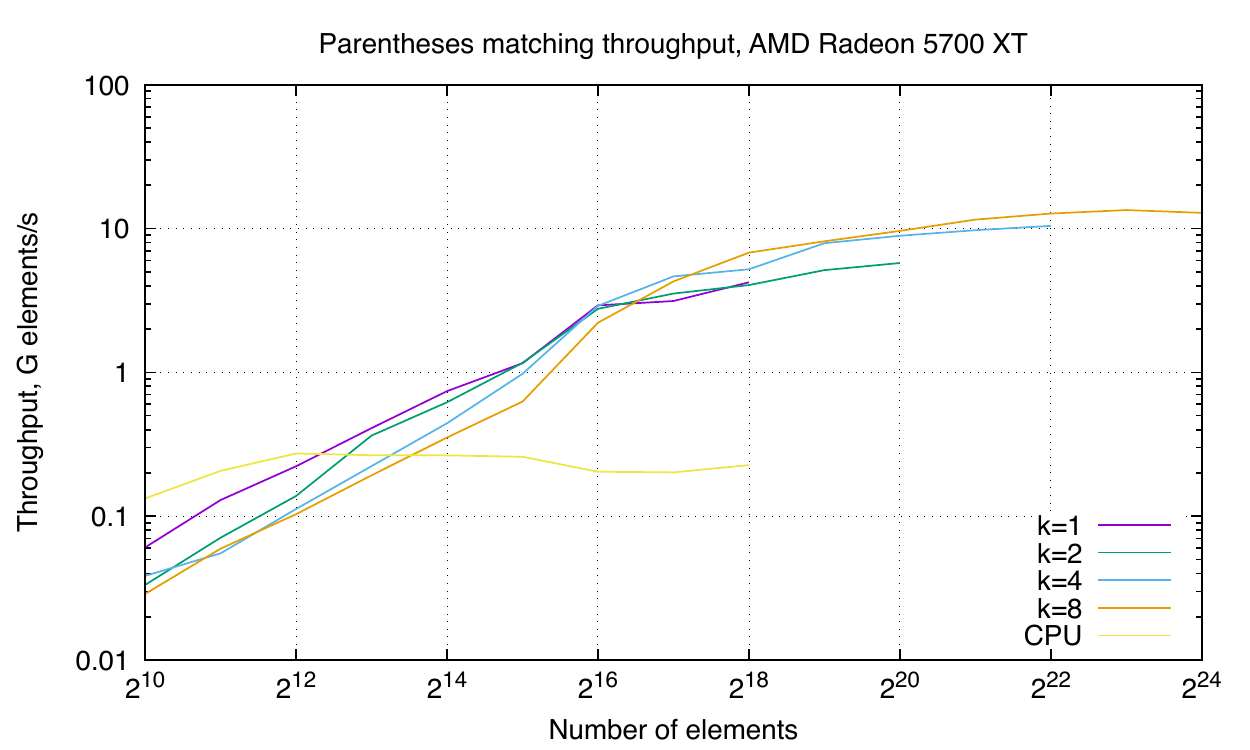}
    \includegraphics[width=190pt]{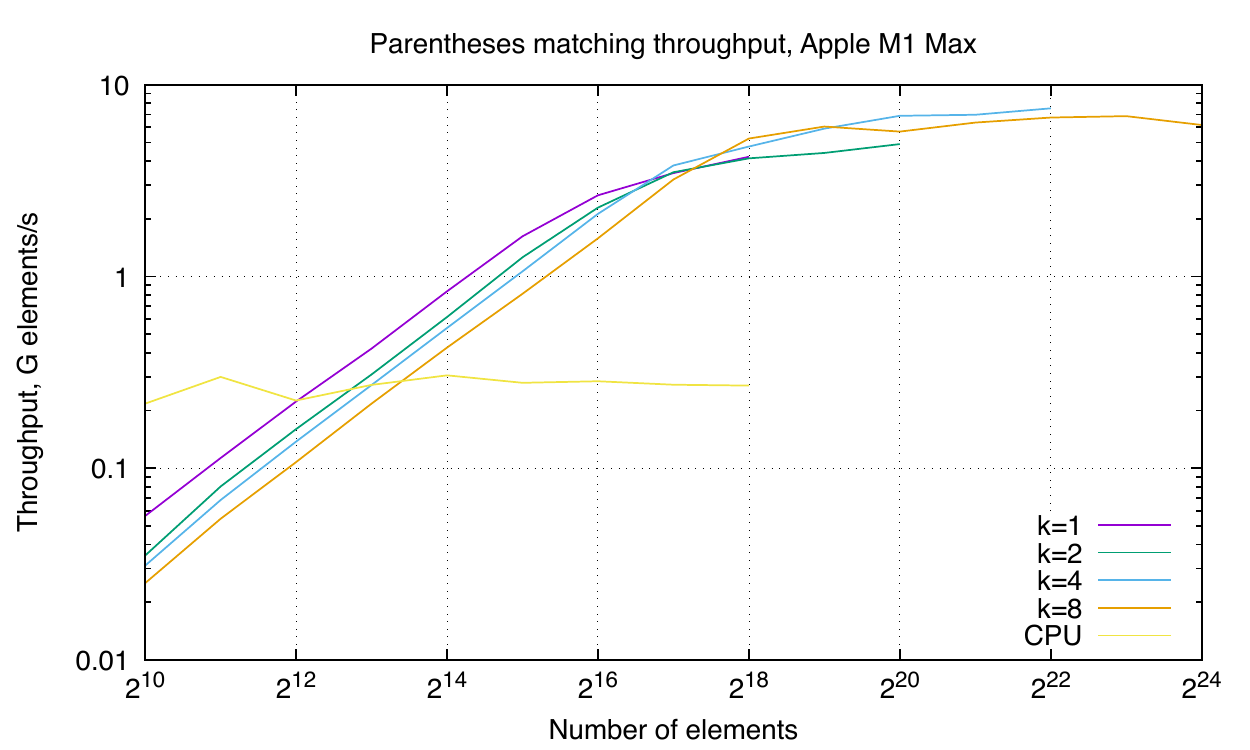}
    \includegraphics[width=190pt]{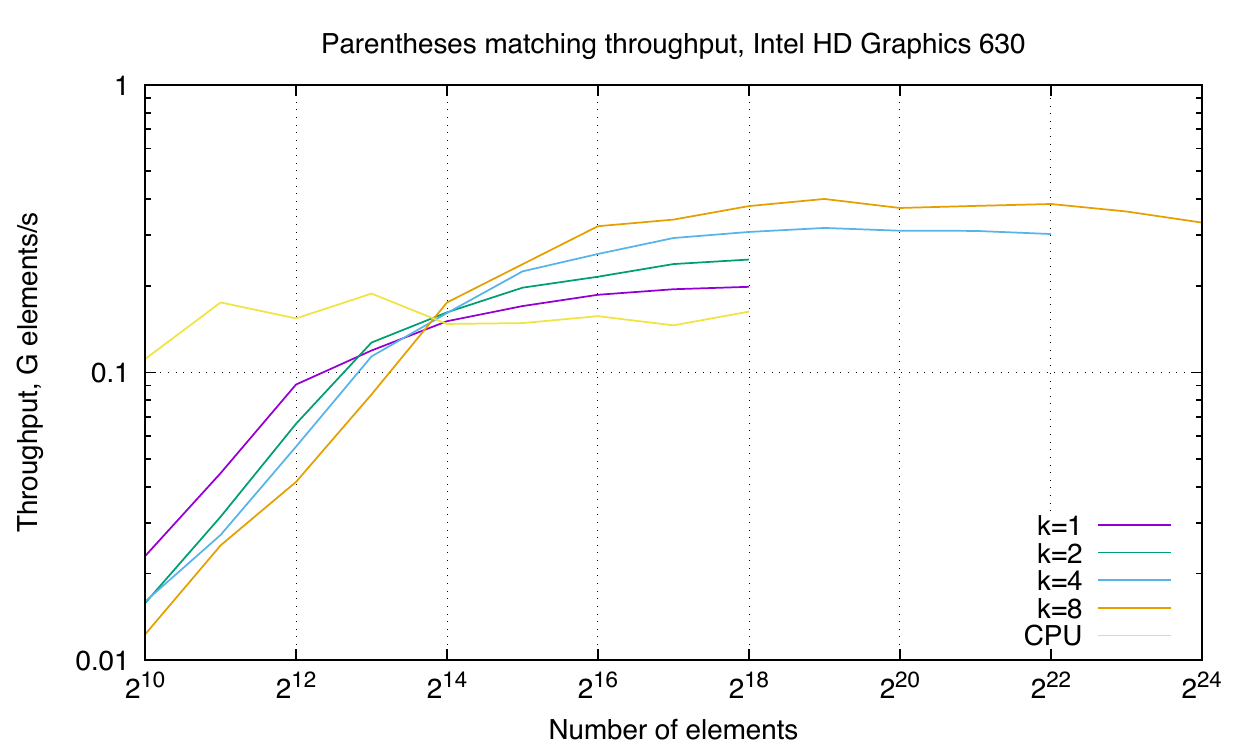}
    \caption{Parentheses matching, AMD 5700 XT, Apple M1 Max, Intel HD 630}
    \label{fig:parenperf}
\end{figure}

The first major performance experiment is to characterize the effect of $k$ on parentheses matching performance, particularly the extent to which increased code complexity of the $k>1$ variant degrades performance. We measured this on three representative GPU devices: AMD Radeon 5700 XT, Apple M1 Max, and Intel HD Graphics 630, each paired with a corresponding CPU (AMD 5950, Apple M1 Max CPU, and i7-4770K, respectively). The results are summarized graphically in Figure~\ref{fig:parenperf}.

There are a number of relevant observations from these measurements. Small problem sizes are dominated by the dispatch overhead, and utilization of GPU processor threads is poor. This is unsurprising, as if there are more processors than elements in the problem, some processors will have no work to do. Even so, we found that the GPU consistently outperformed the CPU for moderate sized to larger inputs; the typical threshold was $2^{12}$.

Similarly, the optimum value for $k$ depends strongly on problem size. The above-mentioned utilization problem becomes worse by a factor of $k$ for small problems, and, in addition, $k>1$ incurs a nontrivial increase in code complexity with a concomitant impact on performance. The $k=1$ implementation is fastest up to and including $2^{16}$ on both tested discrete GPUs, and up to $2^{12}$ on the integrated GPU (the discrepancy is largely explained by the significantly smaller number of processor cores in the latter). For intermediate problem sizes, the tradeoff is very fine, with no decisive advantage for any particular value of $k$. However, for large problem sizes, increasing $k$ decisively improves throughput. Peak throughput is 13.5G element/s (at $k=8$) on the AMD 5700 XT, 7.5G on the Apple M1 (at $k=4$), and 400M on the Intel 630 (at $k=8$). Larger values of $k$ were not observed to improve performance further, almost certainly due to the effects of register pressure, as the number of vector registers required scales linearly with $k$.

The peak performance of 13.5 billion elements/s on the AMD hardware is notable, as bandwidth consumed for simply reading and writing the problem (108GB/s) represents a substantial fraction (27\%) of the raw memory bandwidth of the device (approximately 400GB/s), not counting any processing time for the matching algorithm itself. This device has particularly fast workgroup shared memory operations. Performance of the Intel 630 was less inspiring, also due to the relative performance of workgroup shared memory on that device.
 
Thus, we conclude that the choice of $k$ is dependent on the problem size. Given that the 2D scenes we anticipate are unlikely to have more than a few tens of thousands of clip and blend nodes, we decided that $k=1$ was sufficient for the subsequent implementation work. Applications where larger problems sizes are common may benefit from a different choice. Of course, an adaptive approach where $k$ is chosen in response to the problem size can optimize throughput further, but at nontrivial complexity cost.

\subsection{Bounding box performance}

\begin{figure}
    \includegraphics[width=190pt]{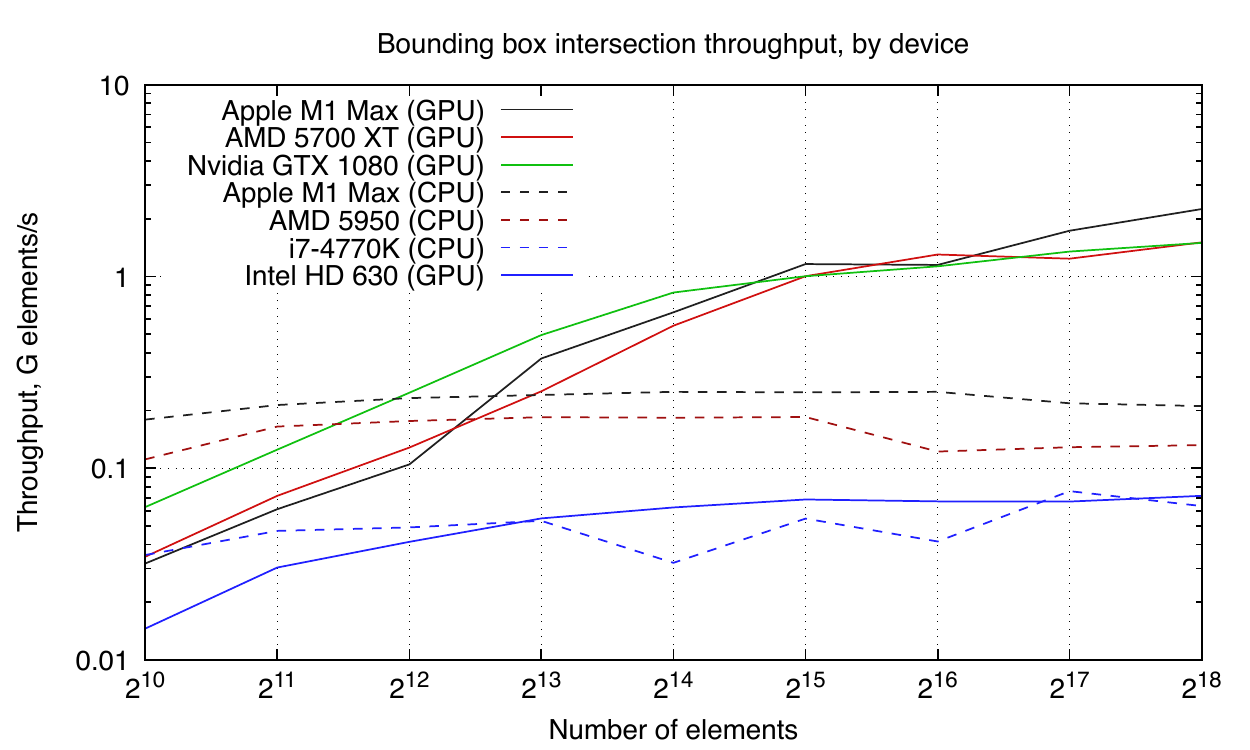}
    \includegraphics[width=190pt]{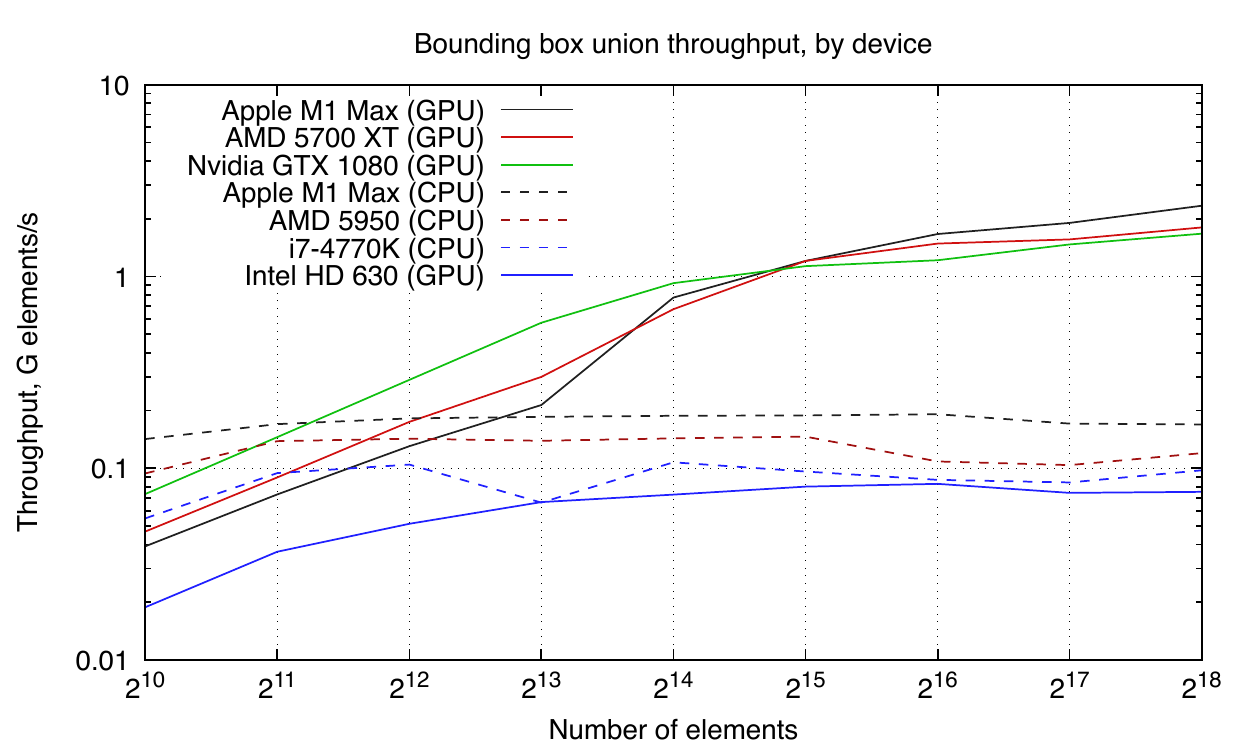}
    \caption{Bounding box union and intersection performance}
    \label{fig:bboxperf}
\end{figure}

While parentheses matching performance informed implementation choices for the rest of the project (and may be interesting for other applications such as parsing), the main result of this paper is performance on the bounding box computation problem. This was measured on the same set of hardware as above, with the addition of an Nvidia GTX 1080. Results are summarized in Figure~\ref{fig:bboxperf}.

The major contours follow similar patterns as parentheses matching. The figure splits out the intersection and union tasks, but since they are so similar algorithmically it is not surprising that performance is similar as well.

On the discrete graphics cards and the M1 Max, the GPU performed the bounding box task with decisively higher performance than the CPU, especially for larger problem sizes; the ratio of peak performance of 5700 XT over 5950 CPU is approximately 12x. Similarly on M1 Max hardware it is approximately 6x (the paired CPU is especially notable for its single-thread performance), and the 1080 is in the same range. Sadly low-end integrated GPU devices such as the Intel 630 do not outperform the CPU on this task, again almost certainly due to poor relative performance of workgroup shared memory. More recent evolutions of the Intel GPU architecture move the location of shared memory from the L3 cache to the subslice, which should significantly improve performance.

\section{Related Work}

There is an extensive literature of algorithms for parentheses matching described in terms of the PRAM model. We will briefly survey those. Generally, an algorithm that runs on $n$ processors in $O(\log n)$ time is straightforward, but adaptations to make it work-efficient add significant complexity.

The first work-efficient algorithm in the literature is \cite{Bar85}. The core of this algorithm is essentially equivalent to the up-sweep of the bicyclic semigroup followed by efficient binary search; they don't describe it in terms of a single semigroup, but rather do two passes, one a simple prefix sum for nesting depth, the second an up-sweep using a minimum operation. Certainly on modern GPUs the bicyclic semigroup formulation is superior, as a single pass is more efficient than two, and the calculation of the semigroup itself compiles to a small number of inexpensive machine operations. The work-efficient adaptation depends on scans in both directions.

Much of the following literature is concerned with efficient execution on weaker PRAM variants, in particular EREW (exclusive read, exclusive write) rather than CREW (concurrent read, exclusive write). These concerns don't map well to actual GPU hardware. Indeed, after a dispatch boundary, having many threads read from the same location is a potentially good for performance, due to caching.

An example of work targeting the EREW model is \cite{Pra94}. While the refinement of the theoretical execution model is of little interest when targeting actual GPU hardware, their Algorithm II is notable because it is effectively an in-place reduction of the stack monoid presented in this paper, though the monoidal structure was not identified as such in that paper.

The parentheses matching problem is very similar that of deriving parent and left sibling vectors from a depth vector. The depth vector is a representation of tree structure popular in the APL world, and it can readily be derived as a prefix sum of {+1/-1} values corresponding to open and close parentheses in the input sequence, respectively. A highly parallel algorithm is given in Section 3.3 of \cite{Hsu19}. This algorithm, however, is not work-efficient, but rather has an additional work factor proportional to the maximum depth of the tree. The present work has no such limitation, and tree depth is unbounded with no impact on performance.

The parentheses matching is related to prefix sum \cite{Ble90}. The latter has both a solid theoretical foundation and a host of practical implementations on actual GPUs, of which a good early example is \cite{Sen08}. We have liberally adapted techniques used for efficient implementation of prefix sum on GPU to develop the algorithm presented here.

Perhaps the most recent relevant work regarding parentheses matching is \cite{Voe21}, an implementation of both a compiler and a JSON parser on modern GPUs in the Futhark language. Its approach to parentheses matching is a fairly straightforward adaption of \cite{Bar85}, and of course they apply significant additional processing for lexical analysis and other tasks related to parsing.

\section{Discussion and future work}

The parentheses matching problem is similar in many ways to prefix sum, for which there is much work on efficient implementations. In particular, both can be represented as monoids, though the monoidal structure of parentheses matching is trickier than pure sums. In particular, for parentheses matching there are two monoids with different time/space tradeoffs, and only through their interleaving is a work-efficient algorithm possible. This algorithm is not merely theoretically work-efficient in a PRAM model, but maps well to efficient implementation on GPU using techniques similar to existing prefix sum implementations.

The main limitation of the proposed algorithm is that the presentation and implementation is a pipeline of 2 dispatches and is limited to inputs of size $w^2k^2$, where $w$ is workgroup size and $k$ is the number of elements processed per thread. In many cases it is possible to increase $k$ to accommodate the problem size (as is the case for the motivating 2D graphics example), but as inputs scale up the algorithm would need to be extended to 3 or more dispatches. A typical 2D graphics scene has at most a few million total nodes, of which a small fraction are clip and blend nodes, so whether such extension to larger problems is needed depends on the problem being solved.

There are a few different approaches for handling larger problem sizes, depending on the exact application. If the nesting depth can be bounded by $wk$, then the most straightfoward approach is a standard tree reduction applied at workgroup granularity; this is work-efficient and straightforward to implement. If unbounded nesting depth is required, other tradeoffs exist.

This work uses a portable subset of the GPU computation model, using workgroup shared memory to communicate between threads within a dispatch, and multiple dispatches to pass data from one workgroup partition to another. The state of the art in high performance prefix sum implementations is decoupled look-back\cite{Mer16}, which performs a single pass rather than multiple dispatches, and uses message passing between workgroups to propagate the partial results. The primary performance advantage of this technique is eliminating the need to read the input twice, which is important because prefix sum is usually limited by raw memory bandwidth. It may be interesting to apply such a single-pass technique to the parentheses matching problem, though it may be tricky, and it is not obvious that memory bandwidth is the limiting factor. Similarly, it is worth exploring whether subgroups can speed the communication of partial results between threads, compared with workgroup shared memory, though with an increase in the difficulty of writing portable compute shader code.

While we've focused on computation of bounding boxes and presented performance data for that problem, the techniques generalize to other applications. While bounding boxes happen to be idempotent and commutative, the algorithm does not depend on those properties. In fact, it can compute any monoid, and it may well be interesting to explore further applications of the techniques, applied to other monoids. 

For both the bounding box problem and other potential applications, this work should help make parsing and other manipulation of tree-structured data practical for implementation on GPU, pushing past the common misconception that this work is inherently serial and must be run on CPU.

\bibliographystyle{ACM-Reference-Format}
\bibliography{stack}

\end{document}